\documentclass[11pt]{elsart}
\usepackage[dvips]{graphicx}

\begin{document}
\begin{frontmatter}

\title{A deterministic small-world network created by edge iterations}
\author{Zhongzhi Zhang\corauthref{zz}}
\corauth[zz]{Corresponding author. Tel. 0086-411-84707819.}
\ead{dlutzzz063@yahoo.com.cn}
\author{Lili Rong}
\ead{llrong@dlut.edu.cn}
\address{
Institute of Systems Engineering, Dalian University of Technology,
\\ Dalian 116024, Liaoning, China}%
\author{Chonghui Guo}
\ead{guochonghui@tsinghua.org.cn}
\address{
Department of Applied Mathematics , Dalian University of Technology,
\\ Dalian 116024, Liaoning, China}%

\begin{abstract}
Small-world networks are ubiquitous in real-life systems. Most
previous models of small-world networks are stochastic. The
randomness makes it more difficult to gain a visual understanding on
how do different nodes of networks interact with each other and is
not appropriate for communication networks that have fixed
interconnections. Here we present a model that generates a
small-world network in a simple deterministic way. Our model has a
discrete exponential degree distribution. We solve the main
characteristics of the
model.\\


\end{abstract}


\date{}
\end{frontmatter}


\section{Introduction}
Small-world networks describe many real-life networks, such as the
World Wide Web, communication networks, the neuronal network of the
worm C. elegans, the electric power grid of southern California, or
social networks that achieve both a strong local clustering and a
small average path length~\cite{AlBa02,DoMe02,Ne03,St01,Ne00}.
Generally, small-world networks are characterized by three main
properties. First, their average path length (APL) doesn't increase
linearly with the system size, but grows logarithmically with the
number of nodes or slower. Second, average node degree of the
network is small. Third, the network has a high average
clustering~\cite{WaSt98} compared to an Erd\"os-R\'enyi (ER) random
network~\cite{ErRe59,ErRe60} of equal size and average node degree.
Average path length can be used to estimate the average transmission
delay. One can always reduce the average path length by adding more
edges. But for some economical and technical reasons, the average
edge number of nodes must be controlled within an acceptable range.
Moreover, the cliquishness tendency of network nodes leads to a
large clustering coefficient.

The first successful attempt to generate networks with high
clustering coefficients and small APL is the model introduced by
Watts and Strogatz (WS model)~\cite{WaSt98}. This pioneering work
of Watts and Strogatz started an avalanche of research on the
properties of small-world networks and the Watts-Strogatz (WS)
model. A much-studied variant of the WS model was proposed by
Newman and Watts~\cite{NeWa99a,NeWa99b}, in which edges are added
between randomly chosen pairs of nodes, but no edges are removed
from the regular lattice. In 1999, Kasturirangan proposed an
alternative model to WS small-world network~\cite{Ka99}. The model
starts also with one ring lattice, then a number of extra nodes
are added in the middle of the lattice which are connected to a
large number of sites chosen randomly on the main lattice. In
fact, even in the case where only one extra node is added, the
model shows the small-world effect if that node is sufficiently
highly connected, which has been solved exactly by Dorogovtsev and
Mendes~\cite{DoMe00}. To investigate the small-world effect
further, Kleinberg has presented a generalization of the WS model
which is based on a two-dimensional square lattice~\cite{Kl00}.
Besides, in order to study other mechanisms for forming
small-world networks, Ozik, Hunt and Ott introduced a simple
evolution model of growing small-world networks, in which all
connections are made locally to geographically nearby
sites~\cite{OzHuOt04}.

The above models are all random. It is known to us all,
stochasticity is a common feature of complex network models that
generate small-world and scale-free topologies. That is, new nodes
connect using a probabilistic rule to the nodes already present in
the system. But as mentioned by Barab\'asi et al., the randomness,
while in line with the major features of real-life networks, makes
it harder to gain a visual understanding of how networks are shaped,
and how do different nodes relate to each other~\cite{BaRaVi01}. In
addition, the probabilistic analysis techniques and random placement
or addition of edges used in most previous studies are not
appropriate for communication networks that have fixed
interconnections, such as neural networks, computer networks,
electronic circuits, and so on. Therefore, it would be not only of
major theoretical interest but also of great practical significance
to construct models that lead to scale-free
networks~\cite{BaRaVi01,IgYa04,DoGoMe02,JuKiKa02,RaBa03,No03,CoFeRa04,AnHeAnSi05,DoMa05,ZhCoFeRo05,ZhWaHuCh04}
and small-world networks~\cite{CoOzPe00,CoSa02} in deterministic
fashions. A strong advantage of deterministic networks is that it is
often possible to compute analytically their properties, for
example, degree distribution, clustering coefficient, average path
length, diameter and adjacency matrix whose eigenvalue spectrum
characterizes the topology.

In this paper, we focus on the small-world network topology
generated in a deterministic way. In 2000, using graph-theoretic
methods Comellas, Oz\'on and Peters introduced deterministic
small-world communication networks~\cite{CoOzPe00}. Two years later,
Comellas and Sampels presented other two deterministic techniques
for small-world networks with constant and variable degree
distributions, respectively~\cite{CoSa02}. Here we propose a simple
construction technique generating a deterministic small-world
network by attaching to edges which was used by Dorogovtsev et al.
to generate pseudofractal scale-free web~\cite{DoGoMe02}. This exact
approach of our model enables one to obtain the analytic solution
for relevant network parameters: degree distribution, clustering
coefficient and diameter. As a result, our network has strong
clustering and small diameter.

\section{The iterative algorithm for the deterministic small-world network}

It is well known that the number of nodes in most of networks in
real world increases exponentially with time. The World Wide Web,
for example, has been increasing in size exponentially from a few
thousand nodes in the early 1990s to hundreds of millions today.
Therefore, our deterministic model is constructed in this
evolutionary way. That is to say, our model is a growing network,
whose size (the number of nodes) increases exponentially with time.

\begin{figure}
\begin{center}
\includegraphics[width=16cm]{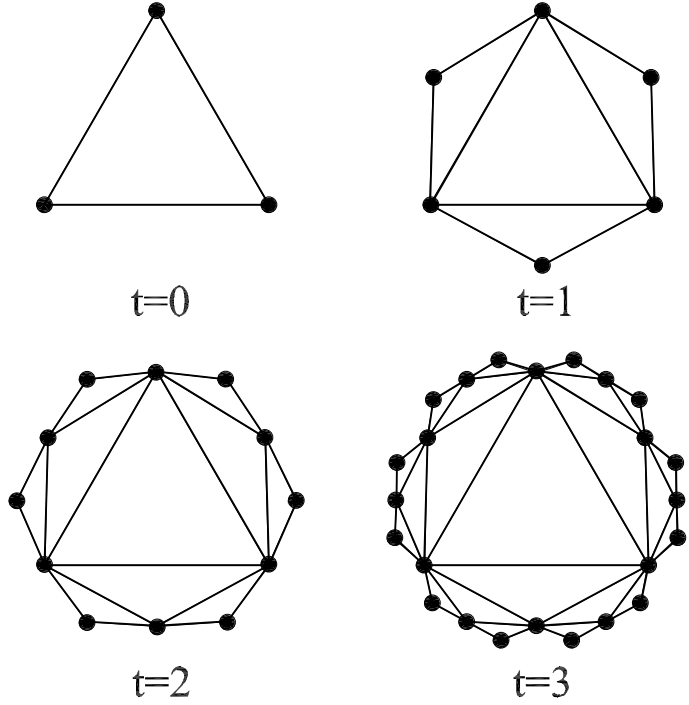}
\caption{Construction of the deterministic small-world network,
showing the first four steps of the iterative process. }
\label{fig:cluster}
\end{center}
\end{figure}

We denote our network after $t$ step evolution by $N(t)$. The
network is created by an iterative method. Its construction
algorithm is the following: For $t=0$, $N(0)$ is a triangle whose
three nodes connect one another. For $t\geq1$, $N(t)$ is obtained
from $N(t-1)$ by adding for each edge created at step $t-1$ a new
node and attaching it to both end nodes of the edge (see Fig. 1).
Notice that in~\cite{DoGoMe02} Dorogovtsev \emph{et al.} introduced
a pseudofractal scale-free web which is constructed as follows: At
each time step, for every edge of the web (not only the ones created
at the last time step as in our network), a new node is added, which
is attached to both end nodes of the edge. So our model is a variant
of the pseudofractal scale-free web. Now we compute the size and
order of $N(t)$. In the evolution process of the model, for each new
node added, two new edges are created in the network. And for each
of the newly-created edges, a node will be created and connected to
both the ends of the edge in the next step. Therefore,
\begin{equation}
\bigtriangleup n_v(t) =n_v(t)-n_v(t-1)=2\bigtriangleup n_v(t-1),
t>1
\end{equation}
where $n_{v}$ is the total number of nodes in the network. Since
$n_{v}(0)=3$, and $\bigtriangleup n_{v}(1)=3$, it follows that
\begin{equation}
\bigtriangleup n_{v}(t) =3\cdot 2^{t-1},  n_v(t) =3\cdot 2^{t}
\end{equation}
The addition of each new node leads to two new edges. Therefore,
\begin{equation}
\bigtriangleup n_e(t) =n_e(t)-n_e(t-1)=2\bigtriangleup
n_v(t)=3\cdot 2^{t}
\end{equation}
where $n_{e}$ is the total number of edges in the network. As
$n_{e}(0)=3$
\begin{equation}
n_e(t) =3\cdot 2^{t+1}-3
\end{equation}
The average node degree is then
\begin{equation}
\langle k\rangle=\frac{2n_{e}}{v_{v}}=\frac{2 (3\cdot
2^{t+1}-3)}{3\cdot 2^{t}}=4\cdot (1-\frac{1}{2^{t+1}})
\end{equation}
For large $t$ it is small and approximately equal to 4. We can see
when $t$ is large enough the resulting network is a sparse graph
whose nodes have many fewer connections than is possible.

\section{Relevant characteristics of the deterministic small-world Network}
Thanks to its deterministic and discrete nature, the model
described above can be solved exactly. In the following we
concentrate on the degree, clustering coefficient and diameter.
\subsection{Degree distribution}
The degree distribution is one of the most important statistical
characteristics of a network. By definition, the degree of a node
$i$ is the number of edges incident from  $i$. We denote the
degree of node $i$  at step $t$ by $k_{i}(t)$. Degree distribution
$P(k)$ is the probability that a randomly selected node has
exactly $k$ edges. By construction, we have
\begin{equation}
k_{i}(t+1)=k_{i}(t)+2
\end{equation}
if $t_{c,i}$ is the step at which a node $i$ is created, then
$k_{i}(t_{c,i})=2$ and hence
\begin{equation}
k_{i}(t)=2(t-t_{c,i}+1)
\end{equation}
Therefore, the degree spectrum of the present network is series of
discrete values: at time $t$, the number of nodes of degree
$k=2\cdot1,2\cdot2,2\cdot3,\cdots,2\cdot $(t-1)$,
2\cdot$t$,2\cdot$(t+1)$ $, equals to $\bigtriangleup
n_{v}(t),\bigtriangleup n_{v}(t-1),\bigtriangleup
n_{v}(t-2),\cdots,\bigtriangleup n_{v}(2),\bigtriangleup
n_{v}(1),n_{v}(0)$, respectively. Other values of degree are absent
in the spectrum. Due to the discreteness of this degree spectrum, it
is convenient to obtain the degree via its cumulative
distribution~\cite{JuKiKa02}, i.e.
\begin{equation}
P(k)=P(k'>k-1)-P(k'>k).
\end{equation}
Using the fact, $P(k'>k)=P\left (t_{c}<\tau=t-(\frac{k}{2}-1)\right
)$, where $t_{c}$ is the birth time of nodes, we obtain that
\begin{equation}
P(k'>k)=\frac{n_{v}(0)}{n_{v}(t)}+\sum_{t_{c}=1}^{\tau-1}\frac{\bigtriangleup
n_{v}(t_{c})}{n_{v}(t)}=2^{-\frac{k}{2}_{.}}
\end{equation}
Using Eq. (8), we have
\begin{equation}
P(k)=P(k'>k-1)-P(k'>k)=(\sqrt{2}-1)2^{-\frac{k}{2}}
\end{equation}
Obviously, when the size of the network is large, the degree
distribution $P(k)$ is an exponential of a power of degree $k$, so
this deterministic small-world model can be called an exponential
network. Note that most small-world networks including WS network
belong to this type ~\cite{BaWe00}.

\subsection{ Clustering coefficient}
Clustering is another important property of a network, which
provides a measure of the local structure within the network. The
most immediate measure of clustering is the clustering coefficient
$C_{i}$ for every node $i$. By definition, clustering
coefficient~\cite{WaSt98} $C_{i}$ of a node $i$ is the ratio of the
total number $E_{i}$ of edges that actually exist between all
$k_{i}$ its nearest neighbors and the number $k_{i}(k_{i}-1)/2$ of
all possible edges between them, i.e.
$C_{i}=2E_{i}/[k_{i}(k_{i}-1)]$. The clustering coefficient $\langle
C \rangle $ of the whole network is the average of all individual
$C_{i}'s$. Next we will compute the clustering coefficient of every
node and their average value.

When a new node $i$ joins the network, its degree $k_{i}$ and
$E_{i}$ is $2$ and $1$, respectively. Each subsequent addition of a
link to that node increases both $k_{i}$ and $E_{i}$ by one. Thus,
$E_{i}$ equals to $k_{i}-1$ for all nodes at all steps. So one can
see that, in this network there is a one-to-one correspondence
between the degree of a node and its clustering. For a node $v$ of
degree $k$, the exact expression for its clustering coefficient is
$2/k$. This expression indicates that the local clustering scales as
$C(k)\sim k^{-1}$. It is interesting to notice that a similar
scaling has been observed empirically in several real-life
networks~\cite{RaBa03}.

Clearly, the number of nodes with clustering coefficient $C=1,
1/2, 1/3, \cdots, $1/(t-1)$, 1/t, $1/(t+1)$ $, is equal to
$\bigtriangleup n_{v}(t), \bigtriangleup n_{v}(t-1),\bigtriangleup
n_{v}(t-2),\cdots,\bigtriangleup n_{v}(2),\bigtriangleup
n_{v}(1),n_{v}(0)$, respectively. The average clustering
coefficient $\langle C \rangle$ can be easily obtained for
arbitrary $t$,
$$
\begin{array}{l}
\langle C\rangle=2\langle\frac{1}{k}\rangle=\frac{1}{n_v(t)}\left[\sum_{i=1}^t\frac{1}{i}\cdot\Delta n_v(t+1-i)+\frac{1}{t+1}\cdot n_v(0)\right]\\
       \quad\quad=1\cdot\frac{1}{2}+\frac{1}{2}\cdot\frac{1}{2^2}+\frac{1}{3}\cdot\frac{1}{2^3}+ \frac{1}{4} \cdot \frac{1}{2^{4}}+\cdots +
\frac{1}{t-1}\cdot \frac{1}{2^{t-1}}+\frac{1}{t}\cdot
\frac{1}{2^{t}}+ \frac{1}{t+1}\cdot \frac{1}{2^{t}} \\
\quad\quad=\sum_{m=1}^t\frac{1}{m}\left(\frac{1}{2}\right)^m+\frac{1}{t+1}\cdot\frac{1}{2^t}
\end{array}
$$

For infinite $t$, $\langle C \rangle =-ln(1-\frac{1}{2})=ln2$ ,
which approaches to a constant value 0.6931, and so the clustering
is high.
\subsection{Diameter}
The small-world concept describes the fact that there is a
relatively short distance between most pairs of nodes in most
real-life networks. The distance between two nodes is the least
number of edges to get from one node to the other. The average path
length is the smallest number of links connecting a pair of nodes,
averaged over all pairs of nodes. The longest shortest path between
all pairs of nodes is called diameter, which is one of the most
important evaluation indexes because it characterizes the maximum
communication delay in the network. Usually those graphs of smaller
node degree and with smaller diameter or average path length are
preferred as interconnection networks. Small diameter is consistent
with the concept of small-world network and it is easier to compute.
So we will study the diameter instead of average path length. Below
we give the precise analytical computation of diameter of $N(t)$
denoted by $Diam(N(t))$.

Clearly, at step $t = 0$ (resp. $t = 1$), the diameter is equal to 1
(resp. 2). At each step $t\geq 2$, one can easily see that the
diameter always lies between a pair of nodes that have just been
created at this step. We will call such newly-created nodes outer
nodes. At any step $t\geq 2$, we note that an outer node cannot be
connected with two or more nodes that were created during the same
step $t^{'}\geq t-1$. Indeed, we know that from step $2$, no outer
node is connected to a node of the initial triangle $N(0)$. Thus,
for any step $t\geq 2$, any outer node is connected with nodes that
appeared at pairwise different steps. Now consider two outer nodes
created at step $t\geq 2$, say $v_{t}$ and $w_{t}$. Then $v_{t}$ is
connected to two nodes, and one of them must have been created
before or during step $t-2$. We repeat this argument, and we end up
with two cases: (1) $t = 2m$ is even. Then, if we make $m$ "jumps",
from $v_{t}$ we arrive in the initial triangle $N(0)$, in which we
can reach any $w_{t}$ by using an edge of $N(0)$ and making $m$
jumps to $w_{t}$ in a similar way. Thus $Diam(N(2m))\leq 2m+1$. (2)
$t = 2m+1$ is odd. In this case we can stop after m jumps at $N(1)$,
for which we know that the diameter is 2, and make $m$ jumps in a
similar way to reach $w_{t}$. Thus $Diam(N(2m+1))\leq 2(m+1)$. It is
easily seen that the bound can be reached by pairs of outer nodes
created at step $t$. More precisely, those two nodes $v_{t}$ and
$w_{t}$ share the property that they are connected to two nodes that
appeared respectively at steps $t-1$, $t-2$.

Hence, formally, $Diam(N(t)) = t+1$ for any $t\geq 0$. Note that
the logarithm value of total number of nodes $N_{t}$ is
approximately equals to $(t+1)ln2$ for large $t$. Thus the
diameter grows logarithmically with the number of nodes and the
average path length increases more slowly than $ln(N_{t})$.

To see why the diameter and average path length grow so slowly, one
can pull the nodes of the networks represented in Fig. 1 to the
circumference of a circle. The older nodes that had once been
nearest neighbors along the circle are pushed apart as new nodes are
inserted into the interval between them. From Fig. 1, we can see
when new nodes are introduced into the system, the original nodes
are not adjacent but, rather, have a great number of new nodes
between them. Thus, growth leads to long-range edges between old
nodes, and these long-range edges similar to the shortcuts in the WS
model~\cite{WaSt98} are responsible for short diameter and average
path length.

Based on the above discussions, our model is a deterministic
small-world network, because it is a sparse one with high clustering
and short diameter and average path length, which satisfy the three
main necessary properties for small-world network.

\section{Conclusion and discussion}
In conclusion, we have presented a simple model that allows one to
construct deterministic small-world networks. We have obtained the
analytic solution for relevant network parameters of the
deterministic model, which are close to those for usual random
small-world
networks~\cite{WaSt98,NeWa99a,NeWa99b,Ka99,DoMe00,Kl00,OzHuOt04}.
 We believe that
our model may help engineers in network topology-designing and
performance-analyzing, it may also help to demystify the small-world
phenomenon. In addition, the model under consideration is a planar
graph which can be drawn on a plane without edges crossing. Many
real-life networks are planar graphs for technical or natural
requirements, such as layout of printed circuits and vein networks
clinging to cutis. We believe our network may help to understand
some properties of real-world planar networks.

\subsection*{Acknowledgment}
This research was supported by the Natural Science Foundation of
China (Grant No. 70431001). The authors are grateful to the
anonymous referees for their valuable comments and suggestions.


\end{document}